\documentclass[12pt,a4paper]{article}
\usepackage[utf8]{inputenc}
\usepackage[T1]{fontenc}
\usepackage{amsmath}
\usepackage{amssymb}
\usepackage{bm}
\usepackage{caption}
\usepackage{subcaption}
\usepackage[title]{appendix}
\usepackage{hyperref}
\usepackage{color}
\usepackage{graphicx}
\usepackage{cite}
\newcommand{\kl}[1]{\textcolor{magenta}{\textbf{KL:} #1}}
 
\begin{document}
	\title{Hydrodynamic Modes of Holographic Weyl Semimetals}
	\author{Nishal Rai$^{1,2}$ and  Karl Landsteiner$^{3}$}
	\date{
		\begin{small}
			$^{1}$ Department of Physics, St. Joseph's College, North Point, Darjeeling, 734104, India. \\
			$^{2}$ Center for Astrophysics Gravitation and Cosmology (CAGC), SRM University Sikkim, Upper Tadong, Sikkim, India. \\
			$^{3}$ Instituto de Fisica Teorica UAM/CSIC, C/ Nicolas Cabrera 13-15, Universidad Autonoma de Madrid, Cantoblanco, 28049 Madrid, Spain  \\
		\end{small}
		\vspace{0.5cm}
	}
	\maketitle
	
	\begin{abstract}
		We study the quasinormal modes of a holographic model of a Weyl semimetal.  The model features quantum phase transition between a topological phase and a trivial phase. We put particular emphasis on the hydrodynamic modes and show that a hydrodynamic mode is present only in the topological Weyl semimetal phase. 
	\end{abstract}

	\section{Introduction}
	The AdS/CFT correspondence  \cite{Maldacena:1997re,Witten:1998qj}, also known as gauge/gravity duality or holographic duality, has been widely applied in studying strongly coupled systems. In particular in condensed matter systems, QCD and hydrodynamic systems \cite{Zaanen:2015oix, Ammon:2015wua,Hartnoll:2018xxg}. In our current study we are interested in an application to a system of interest for condensed matter physics: namely the behavior of a strongly coupled Weyl semimetals. 
	
	In a conventional picture based on electronic quasiparticles, Weyl semimetals are a new class of materials characterised by pointlike singularities in the Brillouin zone at which conduction and valence bands touch \cite{Hosur:2013kxa, Vafek:2013mpa}. In the vicinity of these points, the behaviour of the electronic quasiparticle excitations can be described by left or  right-handed Weyl spinors. Weyl semimetals exhibit robust properties such as chiral anomaly, Fermi arc etc. The physics of Weyl semimetlas has been getting a lot of attentian in the both the experimental and theoretical aspect \cite{Witten:2015aoa,RevModPhys.90.015001}. These Weyl semimetals are also the ideal systems to test transport effects due to quantum anomalies, including chiral magnetic effect \cite{Fukushima:2008xe, Li:2014bha} and transport effects induced by the mixed axial-gravitational anomaly \cite{Gooth_2017}\footnote{See \cite{Chernodub:2021nff} for a theory review.}. The particular interest of Weyl semimetals lies in the fact that they represent a kind of topological quantum matter which cannot be classified using the conventional Landau-Ginzburg classification. We are interested in a holographic description of a strongly coupled Weyl semi-metal without any well defined quasiparticles such as the non-fermi liquid. 
	
	Holography has been used in studying these kind of strongly coupled condensed matter systems \cite{Zaanen:2015oix,Hartnoll:2018xxg,Mukhopadhyay:2017ltk,Mukhopadhyay:2020tky,Atashi:2022ufl}. In \cite{Landsteiner:2015pdh} the authors have have introduced a holographic model of Weyl semimetals, where they have shown that the system shows a quantum phase transition from a  topologically non-trivial state to a trivial one upon the variation of the parameter of the model where the order parameter is for the phase transition was anomalous Hall effect. This model was further extended in \cite{Landsteiner:2015lsa}, where they have  included the finite temperature analysis of the system and found that at finite temperature this quantum phase transition becomes a smooth crossover behaviour. The novel phase structure of these strongly interacting Weyl semimetal was studied in \cite{Liu:2018spp}. In \cite{Ahn:2024ozz} authors have stated that that the chiral magnetic waves (CMW) could be a potential indicator for the chiral anomaly in Weyl semimetals and have investigated along this line. Further studies of holographic models of Weyl semimetals including their Fermi arcs and entanglement entropies can be found in \cite{Copetti:2016ewq, Grignani:2016wyz,Ammon:2016mwa,  Liu:2018bye, Plantz:2018tqf, Ammon:2018wzb, Baggioli:2018afg, Liu:2018djq, Ji:2019pxx, Song:2019asj, Juricic:2020sgg, Baggioli:2020cld, Kiczek:2020ngg, BitaghsirFadafan:2020lkh, Liu:2020ymx, Grandi:2021bsp, Zhao:2021pih, Ji:2021aan, 
Rodgers:2021azg, Zhao:2021qfo, Gao:2023zbd, Baggioli:2023ynu, Bruni:2023ife, Grandi:2023jna, Caceres:2023mqz, Pan:2023kqy, Chu:2024dti, Matsumoto:2024czp }. Holographic models of topological semimetals are reviewed in \cite{Landsteiner:2019kxb}.
	
	In the present work we extend the previous holographic studies of Weyl semimetals by studying the hydrodynamics  of the Weyl semimetals through the investigation of  quasinormal modes (QNMs). 
	QNMs play an important role in understanding the hydrodynamic of a holographic model. Hydrodynamic modes appear as QNMs  in the AdS black hole background whose frequency vanishes in the limit of zero momentum \cite{Policastro:2002se, Kovtun:2005ev}. Our motivation is the following. As we will review, the holographic Wey  semimetal undergoes a quantum phase transition upon varying a dimenionsless ratio of a mass parameter and the background value of a spatial component of an (axial) gauge field. Without the mass parameter the system has an unbroken though anomalous axial symmetry. This axial symmetry is the reason for hydrodynamic bevior of the axial charge, i.e. axial charge shows diffusion and is reflected in the quasinormal mode spectrum by a specific, purely imaginary hydrodynamics mode. Switching on the mass parameter breaks the axial symmetry at the classical level and one therefore would expect that the hydrodynamic mode receives a gap. On the other hand the RG flow is such that the scalar field goes to zero in the far IR in the topological phase and thus one can also argue that the axial symmetry should be restored as an IR symmetry. Then of course a corresponding hydrodynamic mode should be present. In contrast in the topological trivial phase in which the Hall effect is absent it is the axial gauge field in the background that vanishes in the deep IR whereas the scalarfield stays finite. Thus the axial symmetry stays broken all along the (holographic) RG flow and there should not be any hydrodynamic mode. We will see that this is precisely what happens. In order to isolate the relevant physics we will consider the so-called decoupling limit in which the backreaction of the fluctuations on the geometry can be ignored. Then we basically have to study the quasinormal mode spectrum of the gauge field fluctuations which in the case of the axial gauge fields mix with the scalar fields in way very similar to what happens in the holographic superconductor \cite{Amado:2009ts}. The gauge field fluctuations decompose into transverse and longitudinal sectors. Although the hydrodynamic mode sits in the longitudinal one we will also study the transverse sectors and find some interesting behavior as one crosses the quantum phase transition. \\
	We should also clarify that in order to have well-defined quasinormal modes we need to work at finite temperature. Then the transition from the topological to the trivial phase is strictly speaking not a sharp one but rather a smooth cross-over. At sufficiently low temperature this cross-over is however very abrupt and the underlying quantum phase transition at $T=0$ is clearly visible in the Hall conductivity \cite{Landsteiner:2015lsa} and also, as well will show, in the quasinormal mode spectrum.\\
	 We have arranged the article in a following manner, we will begin with the description of the holographic model for Weyl semimetals in section \ref{sec:action}. In section  \ref{sec:background} we will get the numerical solution for the background field where we will consider the probe limit. Followed by section \ref{sec:results} where we will touch upon the method for determining the QNMs from the linear perturbation of the fields.  Then we consider the behaviour of QNMs of the gauge fields in both the transverse and longitudinal sector over the parameter space notably $"M/b"$ and the coefficient $\alpha$ of the Chern-Simon term. In the final section \ref{sec:discussion} we will end with a discussion. 
	
	\section{Action}\label{sec:action}
	We consider the holographic action \cite{Landsteiner:2015lsa} which is given as
	\begin{eqnarray}
		S=\int d^5x \sqrt{-g}\Big[\dfrac{1}{2\kappa^2}(R+12)-\dfrac{1}{4}(F^2+F_5^2)+\nonumber
		\\
		\dfrac{\alpha}{3} \epsilon^{\mu\nu\rho\sigma\tau}A_\mu(F^5_{\mu\rho}F^5_{\sigma\tau}+3F_{\mu\rho}F_{\sigma\tau})-
		\\
		(D_\mu\Phi)^*(D^\mu\Phi)-m^2\Phi^*\Phi\Big],\nonumber
		\label{act}
	\end{eqnarray}
with,
\begin{equation}
F_{\mu\nu}=\partial_\mu V_\nu-\partial_\nu V_\mu,\quad 	F^5_{\mu\nu}=\partial_\mu A_\nu-\partial_\nu A_\mu,\quad D_\mu\Phi=(\partial_\mu-iqA_\mu)\Phi 
\end{equation} 
where $V_\mu$ and $A_\mu$ corresponds to the vector and axial $U(1)$ gauge field respectively and $\Phi$ is a scalar field.  Chern-Simon term $\alpha$ in (\ref{act}) introduced the axial anomaly in the dual field theory. In addition to this, axial symmetry is explicitly broken in case of the axial gauge field through the finite boundary value of the scalar field. We will consider $m^2=-3$ such that the dual operator has dimension three and its source has dimension one. The (consistent) currents for the vector and axial field are expressed as 
\begin{equation}
	\begin{aligned}
		J^\mu=\lim\limits_{u\rightarrow 0}\sqrt{-g}\left(F^{\mu u}+4 \alpha \epsilon^{u\mu\beta\rho\sigma}A_\beta F_{\rho\sigma}\right),\\
		J_5^\mu=\lim\limits_{u\rightarrow 0}\sqrt{-g}\left(F_5^{\mu u}+\dfrac{4}{3} \alpha \epsilon^{u\mu\beta\rho\sigma}A_\beta F^5_{\rho\sigma}\right),
	\end{aligned}
\end{equation}
where 
$u$ is the radial coordinate in the bulk with an asymptotic boundary at $u\rightarrow 0$.

	\section{Background and Ansatz}\label{sec:background}
	We will be working in a probe limit in   Eddington–Finkelstein coordinates, where the ansatz for the metric tensor and the background fields are given as
	\begin{equation}
	\begin{aligned}
			ds^2&=\dfrac{1}{u^2}\left(-f(u)dt^2-2 dtdu+d\vec{x}^2\right),\\
			A&=A_z(u) dz, \quad \Phi = \phi(u),
		\end{aligned}
	\end{equation}
with $f(u)=1-\dfrac{u^4}{u_h^4}$. The horizon lies at $u=u_h$ and the boundary at  $u=0$. This horizon is chosen in such a manner such that $f(u_h)=0$. The temperature is
\begin{equation}
	T=-\dfrac{f'(u_h)}{4\pi}=\dfrac{1}{\pi u_h}.
\end{equation} From onwards we will set the horizon at $u_h=1$, which fixes our temperature.  Introduction of the parameters $b$ and $M$ is done through the boundary terms of $A_z(u)$ and $\phi(u)$ as follows
\begin{equation}
\lim_{u\rightarrow0}A_z(u)=b,\quad \quad \lim_{u\rightarrow0}\dfrac{1}{u}\phi(u)=M,
\end{equation}
where $M$ corresponds to a source for the dual scalar operator that breaks $U(1)$ axial symmetry, $b$ relates to the separation of the chiral nodes \cite{Ahn:2024ozz}.

The	equation of motion for the background fields is given as
	\begin{equation}
		\begin{aligned}
			&A_z''(u)+\left(\frac{f'(u)}{f(u)}-\frac{1}{u}\right) A_z'(u)-\frac{2 q^2 \phi (u)^2 }{u^2
				f(u)}A_z(u)=0,\\
		&	\phi ''(u)+\left(\frac{f'(u)}{f(u)}-\frac{1}{u}\right) \phi '(u)+	\phi (u) \left(-\frac{q^2 A_z(u){}^2}{u^2 f(u)}-\frac{m^2}{u^4
				f(u)}\right)=0.
		\end{aligned}
\label{bceq}
	\end{equation}
We numerically solve the above background equation of motion demanding regularity of the fields at the horizon. In order to do so we will be using a pseudospectral method  \cite{trefethen2000spectral} to get the solution for the background fields where the regularity of the fields near the horizon is naturally satisfied. The background fields are defined as a sum of Chebyshev polynomial in $u$ direction is given by
\begin{equation}
		A_z=\sum_{m=0}^{N-1}A_z^m T_m (2u-1) , \quad \quad
		\phi=\sum_{m=0}^{N-1}\phi^m T_m (2u-1),
	\label{cheb-bac}
\end{equation}
where, $A_z^m$ and $\phi^m$ are the coefficients of the polynomial. These polynomials are inserted into the equation of motion (\ref{bceq})  and the collocation points  for $u=0$ to $u=1$ are chosen in Gauss-Lobatto grid. By choosing the appropriate number of grid points corresponding to the number of coefficients one can solve the set of algebraic equations in terms of these coefficients to get the solution for the background fields using (\ref{cheb-bac}).

\subsection{Numerical solution of the background}
To solve the equation of motion (\ref{bceq}) numerically  for the background field $\phi(u)$ and $A_z(u)$, we will fix  $q=1$,  $b=11$ and vary the dimensionless quantity $M/b$, which one can do due to the underlying conformal symmetry as discussed in \cite{Landsteiner:2015lsa}.
 We are interested in a low temperature limit at which  the system undergoes a
quantum phase transition from a topologically non-trivial to a trivial state, so the temperature is set at $\pi T/b=1/11$.

\begin{figure}[t]
	\centering
	\begin{subfigure}{0.45\textwidth}
		\centering
		\includegraphics[width=1\linewidth]{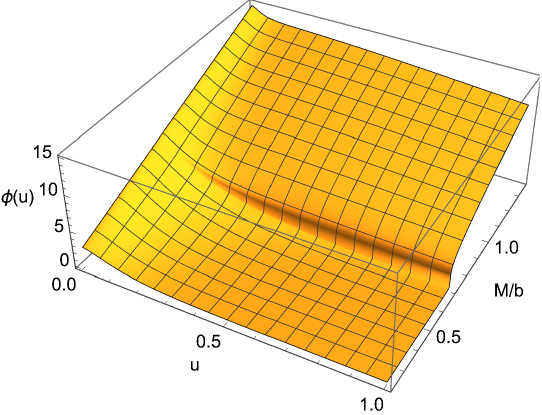}
	\end{subfigure}\hfill
	\begin{subfigure}{0.45\textwidth}
		\includegraphics[width=1\linewidth]{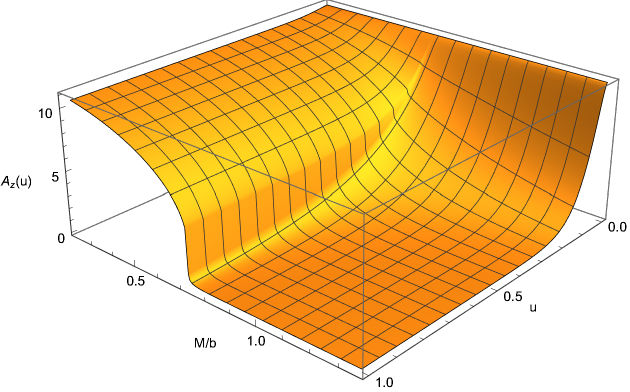}
	\end{subfigure}
	\caption{Plot of $\phi(u)$ (Left) and  $A_z$ (Right) vs $u$ over the parameter space of $M/b$.}
	\label{bc3d}
\end{figure}

\begin{figure}[t]
	\centering
	\includegraphics[width=0.45\linewidth]{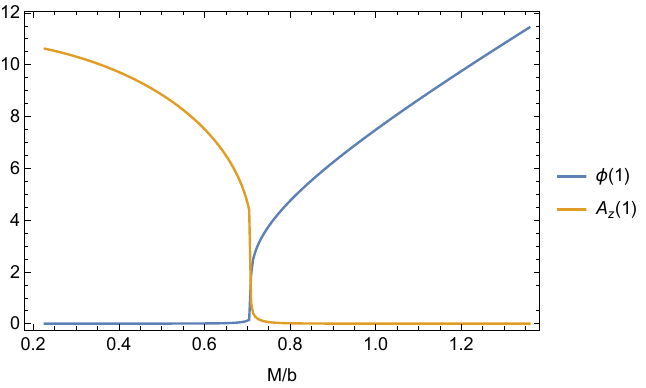}
	\caption{Background field at near horizon.}
	\label{fluh}
\end{figure}

The solutions for the background fields are given in Fig.\;\ref{bc3d}, where we have plotted $\phi(u)$ and $A_z(u)$ vs $u$ over the parameter space of $M/b$. One can see that there is a critical point around $M/b\approx 0.71$, where the system goes from a topological phase ($b>>M$) to a trivial phase. This is more visible in Fig. \ref{fluh}, where we have plotted the value of $\phi(u)$ and $A_z(u)$ at near horizon limit vs $M/b$. Here one can see that as the value of $M/b$ reaches the critical point $M/b\approx 0.71$ \cite{Landsteiner:2015pdh} there is a sharp fall in $A_z (u_h)$ and sharp rise in $\phi (u_h)$.  The choice of the value $b/\pi T$ is determined basically by the numerics. To reach even lower temperatures increased numerical accuracy would be necessary. At the low temperature we are working there is no sharp quantum phase transition but the corresponding cross over is indeed extremely sharp. Thus this choice of parameters is optimally suited for our purpose.


\section{Results}\label{sec:results}
To compute the quasinormal modes, we will use the linear perturbations in the fields. The fields are defined in terms of the background fields and fluctuations as follows
\begin{equation}
	\begin{aligned}
	\Phi&=\phi(u)+\sigma(u,t,z)+i \eta (u,t,z)\\
	A_\mu&=A_z(u)+a_\mu(u,t,z)\\
	V_\mu&=v_\mu(u,t,z).
	\end{aligned}
\end{equation}
A general procedure of Fourier decomposition is used to write the equation of motion for the fluctuations in terms of $\omega$ and $k$. These are  dimensionless quantities and they are related to the physical ones ($\omega_p, k_p$) as 
\begin{equation}
	\omega=\dfrac{\omega_p}{\pi T} \quad \text{and}\quad  k=\dfrac{k_p}{\pi T}.
\end{equation}
To get the quasinormal modes we will be using pseudospectral method with the fluctuations being expressed as follows
\begin{equation}
		\begin{aligned}
		\sigma&=\sum_{m=0}^{N-1}\sigma^m T_m (2u-1) , \quad \quad
		\eta=\sum_{m=0}^{N-1}\eta^m T_m (2u-1), \\
		a_\mu&=\sum_{m=0}^{N-1}a_\mu^m T_m (2u-1), \quad \quad
		v_\mu=\sum_{m=0}^{N-1}v_\mu^m T_m (2u-1) .
	\end{aligned}
\label{cheb}
\end{equation}
As the perturbation is linear in nature one can write the whole equation of motion in a matrix form such that the coefficients in (\ref{cheb}) becomes the column vector with the first and last element being the value of the fluctuations at the horizon and the boundary respectively. The derivatives turn into a differential matrix operator and the whole problem turn into a generalized eigen value problem, where the eigen values corresponds to the value of $\omega$ or the position of the QNMs in a complex plane.

We will present the result in the following way, first we will begin with the study of the QNMs for the transverse sector with the perturbation in both the axial and vector field followed by the similar procedure in the longitudinal sector. 
\subsection{Transverse sector}
\subsubsection{Vector field}
First we will start with the a perturbation in $v_\mu$ with $\mu=(x$, $y$). 
The  equation of motion for the fluctuations is given as
\begin{equation}
	\begin{aligned}
		v_x''(u)+ \left(\frac{f'(u)}{f(u)}+\frac{2 i \omega}{u^2 f(u)}+\frac{1}{u}\right)v_x'(u)-
		&\left(\frac{k^2}{u^2 f(u)}+\frac{i \omega}{u^3
			f(u)}\right)v_x(u)\\
		&+\frac{8 i \alpha  \omega  A_z'(u)}{u f(u)}v_y(u)=0,\\
		v_y''(u)+ \left(\frac{f'(u)}{f(u)}+\frac{2 i \omega}{u^2 f(u)}+\frac{1}{u}\right)v_y'(u)-
		&\left(\frac{k^2}{u^2 f(u)}+\frac{i \omega}{u^3
			f(u)}\right)v_y(u)\\
		&-\frac{8 i \alpha  \omega A_z'(u)}{u f(u)}v_x(u) =0.
	\end{aligned}
	\label{tr-vec}
\end{equation}
\begin{figure}
	\centering
	\begin{subfigure}{0.33\textwidth}
		\includegraphics[width=1\linewidth]{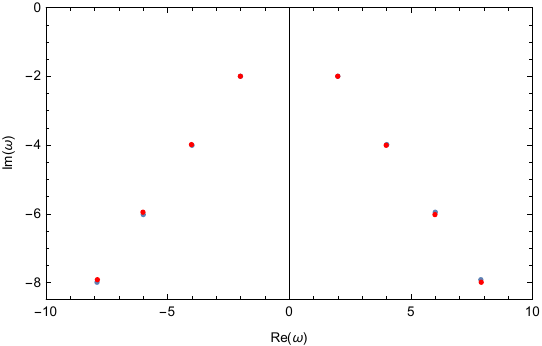}
	\end{subfigure}\hfill
	\begin{subfigure}{0.33\textwidth}
		\includegraphics[width=1\linewidth]{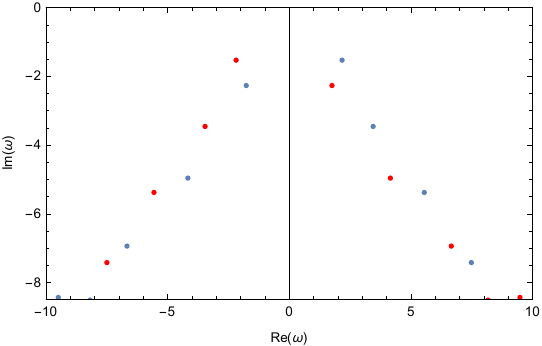}
	\end{subfigure}\hfill
	\begin{subfigure}{0.33\textwidth}
		\includegraphics[width=1\linewidth]{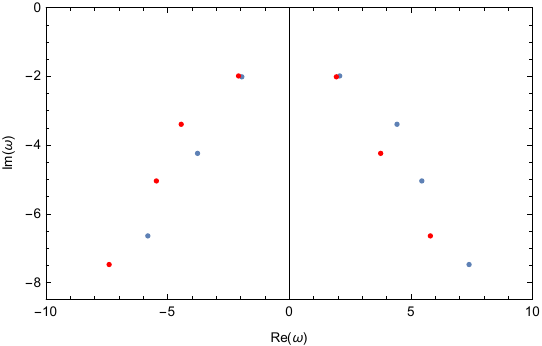}
	\end{subfigure}\hfill
	\caption{QNM for $M/b=0.227$(Left),  $M/b=0.7$ (Middle) and $M/b=1.36$(Bottom-Right). Blue and Red dot corresponds to the $v_{\pm}=v_x\pm v_y$ respectively.}
	\label{disp7}
\end{figure}

\begin{figure}[h]
	\centering
	\begin{subfigure}{0.33\textwidth}
		\centering
		\includegraphics[width=1\linewidth]{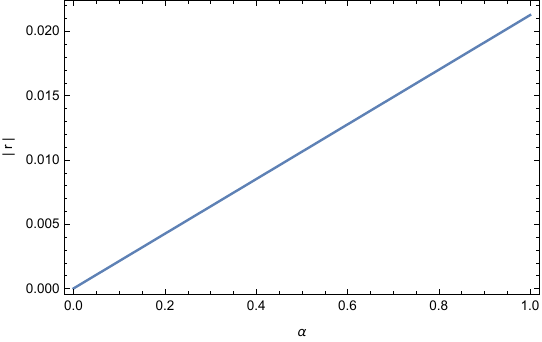}
	\end{subfigure}\hfill
	\begin{subfigure}{0.33\textwidth}
		\includegraphics[width=1\linewidth]{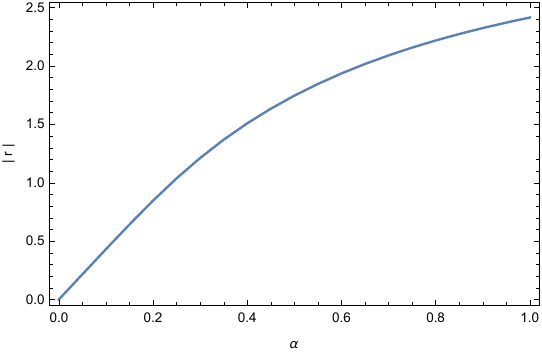}
	\end{subfigure}
	\begin{subfigure}{0.33\textwidth}
		\includegraphics[width=1\linewidth]{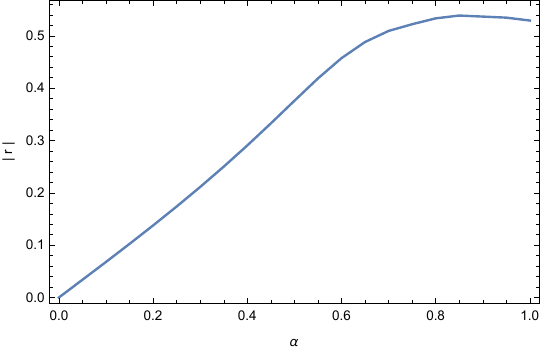}
	\end{subfigure}
	\caption{Plot of separation ($r$)  vs $\alpha$ for the lowest mode (Left)$M/b=0.227$, $M/b=0.7$ (Middle) and $M/b=1.36$ (Right). For vector field - Transverse.}
	\label{fig:sepv}
\end{figure}

One can write these set of coupled differential equation in terms of $v_{\pm}=v_x\pm i v_y$ which decouples them in terms of $v_\pm$. 
We will solve these decoupled differential equations using the pseudospectral method as stated in the previous section. We have plotted Re$(\omega)$ vs Im$(\omega)$ in Fig. \ref{disp7}, for different values of $M/b$ with $\alpha=0.2$. From the left figure one can see that the spectrum is very similar to the one in \cite{Kovtun:2005ev}\footnote{The quasinormal modes of the transverse sector are depicted in the right panel of fig. 4 there.}. This is due to the fact at the range of $M/b<< 0.71$ i.e below critical point the background field $A_z$ is almost constant. In this case the contribution to the Chern-Simons term in (\ref{tr-vec}) is very small from the derivative of $A_z$.
At the limiting case i.e. $M/b=0$ it is constant and there is no contribution from the Chern-Simon term as one can see from (\ref{tr-vec}) and one does not expect any kind of distinction between the two field $v_x$ and $v_y$. Following the similar line of thought one can see that spectrum for $v_x$ and $v_y$ are spread out from each other around the critical region.  

\begin{figure}
	\centering
	\begin{subfigure}{0.35\textwidth}
		\includegraphics[width=1\linewidth]{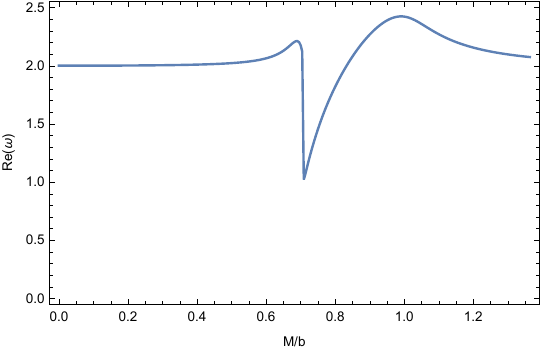}
	\end{subfigure}\hskip2cm
	\begin{subfigure}{0.35\textwidth}
		\includegraphics[width=1\linewidth]{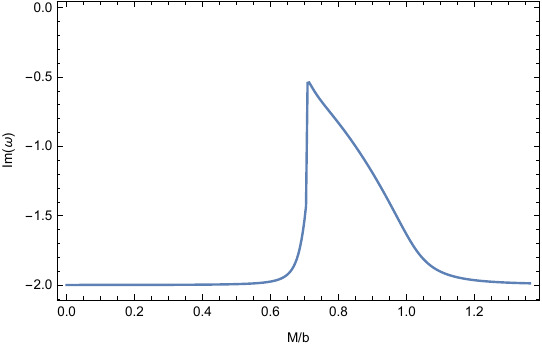}
	\end{subfigure}
	\caption{Plot of Real and complex part of $\omega$ vs $M/b$ for the first lowest mode with $k=0$. (Vector field - Transverse)}
	\label{pwmb}
\end{figure}
The Chern-Simon term lifts the degeneracy in the spectrum of QNMs. To make it more concrete we have plotted the absolute value of the separation $r$ of the lowest QNMs  of $v_x$ and $v_y$ \footnote{The distance $r$ can be evaluated by considering the distance between the lowest modes. Upon considering the $v_\pm$ linear combinations one can simply take the distance between the lowest and the second lowest mode i.e. the distance between lowest red and blue dot either in positive or negative $Re(\omega)$ axis in Fig. \ref{fig:sepv}} in  Fig. \ref{fig:sepv} where one can see that with the increase in $\alpha$ the separation increases. Below the critical point the increase is linear in nature, around the critical point the magnitude of the increase in large and later it becomes less and finally above the critical point the behaviour is very similar to the one near the critical point but the magnitude is much less as compared to critical one. For $\alpha\rightarrow0$ one may approximate the behaviour of the separation as linear with the gradient of the separation being largest near the critical point.  In addition to this we have also plotted how the position of the lowest mode changes with the variation of $M/b$ in Fig. \ref{pwmb} with $k=0$, where one can see that the the real and the complex part of $\omega$ starts at the same initial value at $M/b=0$ and moves toward the same initial value for $M>>b$.
\subsubsection{Axial field}
\begin{figure}
	\centering
	\begin{subfigure}{0.33\textwidth}
		\includegraphics[width=1\linewidth]{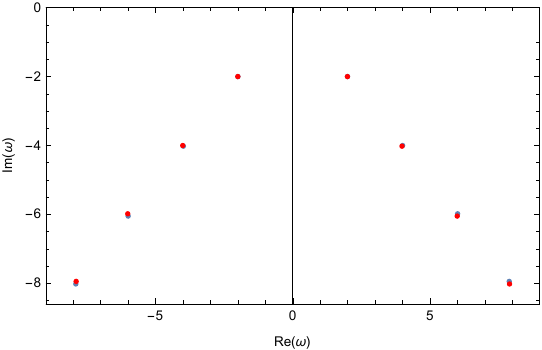}
	\end{subfigure}\hfill
	\begin{subfigure}{0.33\textwidth}
		\includegraphics[width=1\linewidth]{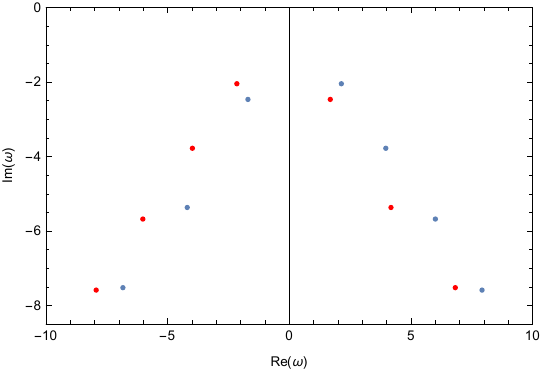}
	\end{subfigure}\hfill
	\begin{subfigure}{0.33\textwidth}
		\includegraphics[width=1\linewidth]{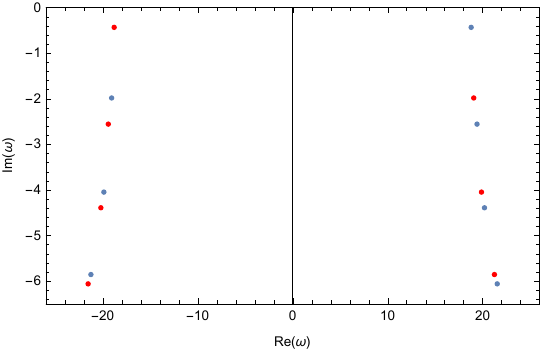}
	\end{subfigure}\hfill
	\caption{QNM for $M/b=0.227$(Left),  $M/b=0.7$ (Middle) and $M/b=1.36$(Bottom-Right). Blue and Red dot corresponds to the $a_{\pm}=a_x\pm a_y$ respectively.}
	\label{disp7b}
\end{figure}

\begin{figure}
	\centering
	\begin{subfigure}{0.33\textwidth}
		\centering
		\includegraphics[width=1\linewidth]{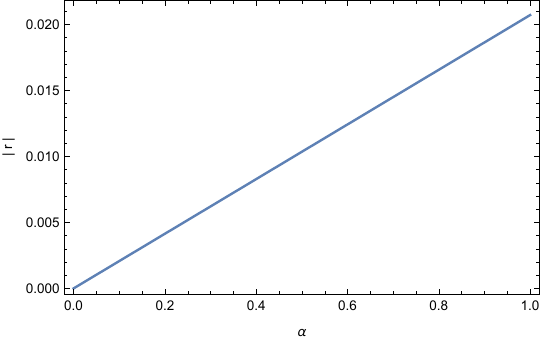}
	\end{subfigure}\hfill
	\begin{subfigure}{0.33\textwidth}
		\includegraphics[width=1\linewidth]{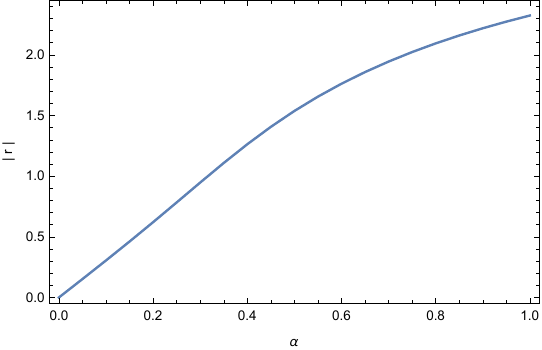}
	\end{subfigure}
	\begin{subfigure}{0.33\textwidth}
		\includegraphics[width=1\linewidth]{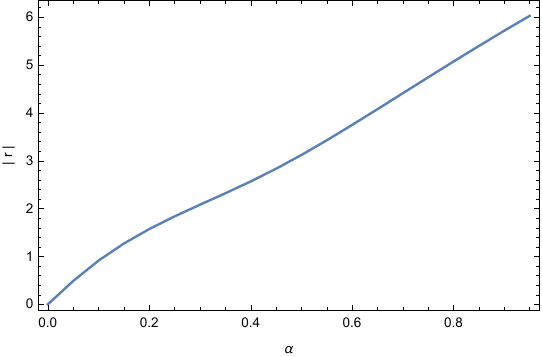}
	\end{subfigure}
	\caption{Plot of separation ($r$)  vs $\alpha$ for the lowest mode (Left)$M/b=0.227$, $M/b=0.7$ (Middle) and $M/b=1.36$ (Right). For Axial field - Transverse.}
	\label{fig:sep}
\end{figure}

\begin{figure}
	\centering
	\begin{subfigure}{0.35\textwidth}
		\includegraphics[width=1\linewidth]{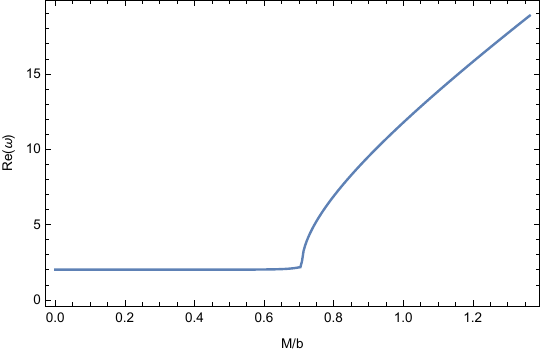}
	\end{subfigure}\hskip2cm
	\begin{subfigure}{0.35\textwidth}
		\includegraphics[width=1\linewidth]{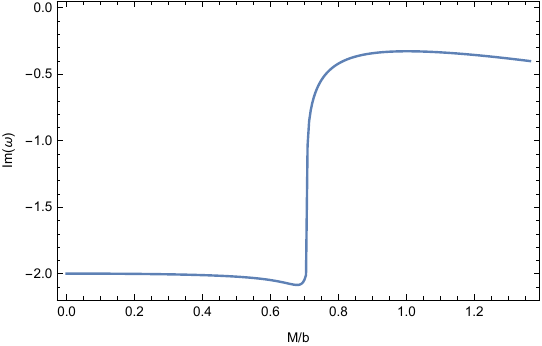}
	\end{subfigure}
	\caption{Plot of Real and complex part of $\omega$ vs $M/b$ for the first lowest mode with $k=0$.(Axial field - Transverse)}
	\label{axmb}
\end{figure}

The equation of motion for the fluctuation of the axial field is given by 
\begin{equation}
	\begin{aligned}
		a_x''(u)+ \left(\frac{f'(u)}{f(u)}+\frac{2 i \omega}{u^2
			f(u)}+\frac{1}{u}\right)a_x'(u)-\frac{8 i \alpha  \omega a_y(u)
			A_z'(u)}{u f(u)}\\
		- \left(\frac{k^2}{u^2 f(u)}+\frac{2 q^2
			\phi (u)^2}{u^4 f(u)}+\frac{i \omega}{u^3 f(u)}\right)a_x(u)
		=0,\\
		a_y''(u)+ \left(\frac{f'(u)}{f(u)}+\frac{2 i \omega}{u^2
			f(u)}+\frac{1}{u}\right)a_y'(u)+\frac{8 i \alpha  \omega a_x(u)
			A_z'(u)}{u f(u)}\\
		- \left(\frac{k^2}{u^2 f(u)}+\frac{2 q^2
			\phi (u)^2}{u^4 f(u)}+\frac{i \omega}{u^3 f(u)}\right)a_y(u)
		=0.
	\end{aligned}
	\label{tr-axl}
\end{equation}
Here the above equations can be diagonalised using $a_\pm=a_x\pm i a_y$, similar to the previous case. 
We have set the values of the parameters as follows: $\alpha=0.2$ and $q=1$. 
Later the variation of $\alpha$ is considered for different values of $M/b$. 
The behaviour of the QNMs are very similar to the one for the perturbation of the vector field expect for the fact that the modes will try to move away from their initial value of $M/b\rightarrow0$ after the transition point $M/b\approx 0.71$. In Fig. \ref{disp7b} we have given the QNMs where one can see that after the transition point the modes moves away from the initial value of $M/b\rightarrow0$, this is due to the fact that unlike the perturbation of the vector field in the axial field perturbation there is a contribution from the background scalar field. This behaviour is clearer in Fig. \ref{fig:sep} where we have plotted absolute value of the separation $r$ vs $M/b$, where for the $M/b=1.36$ i.e. above the critical point the separation keeps on increasing. In  Fig. \ref{axmb} we have plotted the behaviour of the lowest mode with respect to $M/b$ where we can see that the mode moves away from its initial value of $M/b=0$ for $M>b$.


\section{Longitudinal Sector}
\subsection{Vector field}
We have considered the fluctuation along the $z$ and $t$ direction for the vector field $v_\mu$. 
In this case, none of the background field i.e. $\phi$ and $A_z$ nor the Chern-Simon coupling $\alpha$ enter the equation of motion. Hence, it will turn into the system discussed in \cite{Kovtun:2005ev} 
	, which gives a diffusive mode.
The equation of motion is given by 
\begin{equation}
	\begin{aligned}
		&v_t''(u)-\frac{a_t'(u)}{u}-i k v_z'(u)=0,\\
		&v_z''(u)+ \left(\frac{f'(u)}{f(u)}+\frac{2 i \omega}{u^2 f(u)}+\frac{1}{u}\right)v_z'(u)-\\ 
		&\frac{i}{u^3 f(u)} \bigl(k (v_t(u)- u v_t'(u))+\omega v_z(u)\bigl)=0,\\
		&\omega a_t'(u)+u^2 f(u) k a_z'(u)+i k \left(k a_t(u)+\omega a_z(u)\right)=0.
	\end{aligned}
\end{equation}

\begin{figure}[b]
	\centering
	\begin{subfigure}{0.35\textwidth}
		\centering
		\includegraphics[width=1\linewidth]{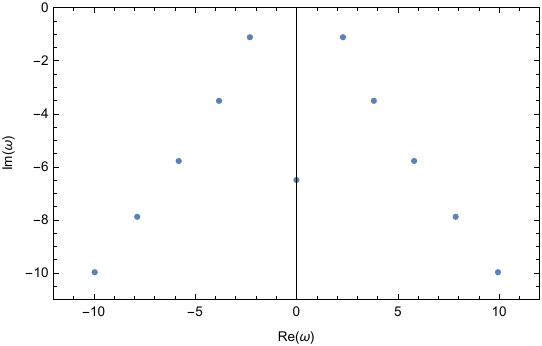}
	\end{subfigure}\hskip2cm
	\begin{subfigure}{0.35\textwidth}
		\includegraphics[width=1\linewidth]{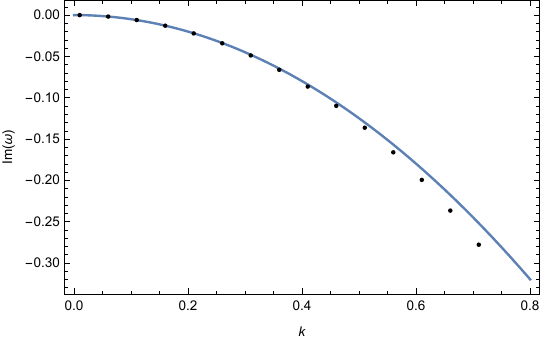}
	\end{subfigure}
	\caption{Plot of Im$(\omega)$  vs Re$(\omega)$ for $k=2$ (Left) and  Im$(\omega)$ vs $k$ for the diffusive mode with solid blue line corresponds to  $\omega=-i \; 0.5 k^2$ (Right). (Lowest mode) (Vector-Longitudinal)}
	\label{diff}
\end{figure}
The above first two equations together with the constraint equation in the last line  can be written in terms of the gauge invariant term $E_z=k v_t+\omega v_z$. The dispersion relation and  spectrum of the mode is given in 	Fig. \ref{diff}. We have not taken the variation of the other parameters of the system here as this modes or the equation of motion of these fields are independent of the background fields.

\subsection{Axial field}
Finally we will consider the  the fluctuation of the axial field $a_\mu$ along the $z$ and $t$ direction for the vector field. Equation of motion for the fluctuations of the fields is given by
\begin{equation}
	\begin{aligned}
		&a_t''(u)	-\frac{a_t'(u)}{u}+2 q \left( \sigma (u) \phi '(u)- \phi (u) \sigma '(u)\right)-i k A_z'(u)=0,\\ 
		&a_z''(u)+a_z'(u) \left(\frac{f'(u)}{f(u)}+\frac{2 i \omega}{u^2 f(u)}+\frac{1}{u}\right)-a_z(u) \left(\frac{2 q^2 \phi (u)^2}{u^2 f(u)}+\frac{i \omega}{u^3 f(u)}\right)+\\
		&\frac{i k}{u^3 f(u)}(u a_t'(u)-a_t(u))+\dfrac{2 i q }{u^2 f(u)} \left(k \eta (u)+2 i q A_z(u) \sigma (u)\right)\phi (u)=0,\\
		&\sigma ''(u)+ \left(\frac{f'(u)}{f(u)}+\frac{2 i \omega}{u^2
			f(u)}+\frac{1}{u}\right)\sigma '(u)-\frac{2 q }{u^2 f(u)}(q \phi (u) a_z(u)-i k \eta (u)) A_z(u) +\\
		&\frac{1}{u} \left(-\frac{q^2 A_z(u){}^2}{u f(u)}+\frac{f'(u)}{f(u)}-\frac{k^2}{u
			f(u)}-\frac{m^2}{u^3 f(u)}-\frac{i \omega}{u^2 f(u)}-\frac{1}{u}\right)\sigma (u)=0,\\
		&\eta ''(u)+\left(\frac{f'(u)}{f(u)}+\frac{2 i \omega}{u^2 f(u)}+\frac{1}{u}\right)\eta '(u) -\frac{i k q }{u^2 f(u)} \left(\phi (u) a_z(u)+2 \sigma (u) A_z(u)\right)+\\
		&\dfrac{1}{u}\left(-\frac{q^2 A_z(u){}^2}{u f(u)}+\frac{f'(u)}{f(u)}-\frac{k^2}{u
			f(u)}-\frac{m^2}{u^3 f(u)}-\frac{i \omega}{u^2 f(u)}-\frac{1}{u}\right)\eta (u)+\\
		&\frac{q}{u^3 f(u)}(2 u a_t(u) \phi '(u)+\phi (u) \left(u a_t'(u)-a_t(u)\right))=0,\\
		&k a_z'(u)+2 i q \phi (u) \eta
		'(u)+\frac{\omega a_t'(u)}{u^2 f(u)}+\frac{i}{u^2 f(u)} \left(k^2+2 q^2 \phi (u)^2\right)a_t(u)-\\
		& q \left(\frac{2 \omega \phi (u)}{u^2 f(u)}-2 i \phi '(u)\right) \eta (u)+\frac{i k \omega a_z(u)}{u^2 f(u)}=0
	\end{aligned}
\end{equation}

where, the last equation is the constraint equation. The equation of motion are  invariant under following gauge transformation
	\begin{eqnarray*}
	\delta a_\mu&=&\partial_\mu\lambda ,\\
\delta \sigma &=& -q\; \phi ,\\
\delta \eta&=&0 .
\end{eqnarray*}
The above equations can also be rewritten in terms of the gauge invariant quantities $\Sigma (u), E_z (u)$ and $\eta(u)$ which are defined as follows
\begin{eqnarray*}
	E_z(u)&=&k a_t(u)+\omega a_z(u),\\
	\Sigma(u)&=&(\omega-k) \sigma (u)+i q \phi (u) \left(a_t(u)+a_z(u)\right),\\
	\eta(u)&=&\eta(u) .
\end{eqnarray*}
\begin{figure}[t]
	\centering
	\includegraphics[width=0.45\linewidth]{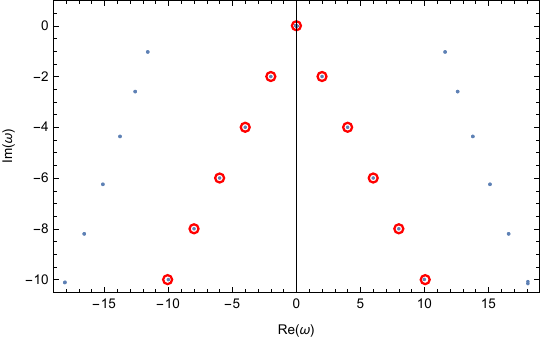}
	\caption{Plot of Im$(\omega)$  vs Re$(\omega)$ for M/b=0 with $k=0.1$. The red circles are the solution for the vector type perturbation which coincides with $E_z(u)$ of the axial at the limit of $M/b\rightarrow0$. }
	\label{fig:long-qnm}
\end{figure}

\begin{figure}[h]
	\centering
	\begin{subfigure}{0.33\textwidth}
		\centering
		\includegraphics[width=1\linewidth]{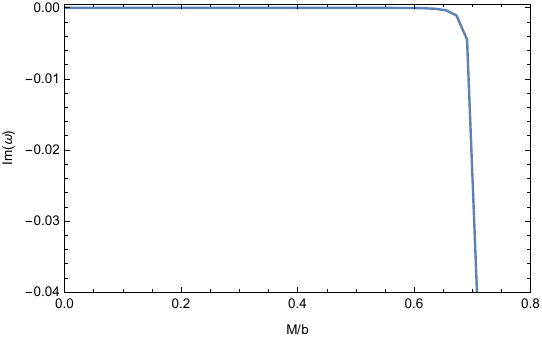}
	\end{subfigure}\hfill
	\begin{subfigure}{0.33\textwidth}
		\centering
		\includegraphics[width=1\linewidth]{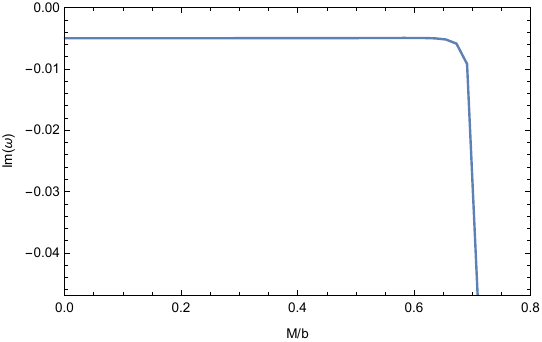}
	\end{subfigure}\hfill
	\begin{subfigure}{0.33\textwidth}
		\includegraphics[width=1\linewidth]{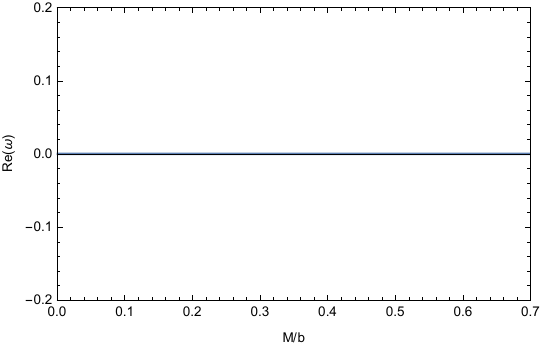}
	\end{subfigure}
	\caption{Plot of Im$(\omega)$  for $k=0.001$ (Left), $k=0.1$(Middle)  and Re$(\omega)$ vs M/b for $k=0.1$, $0.001$ (Right)  . (For the diffusive mode)}
	\label{long}
\end{figure}

\begin{figure}[h]
	\centering
	\begin{subfigure}{0.35\textwidth}
		\centering
		\includegraphics[width=1\linewidth]{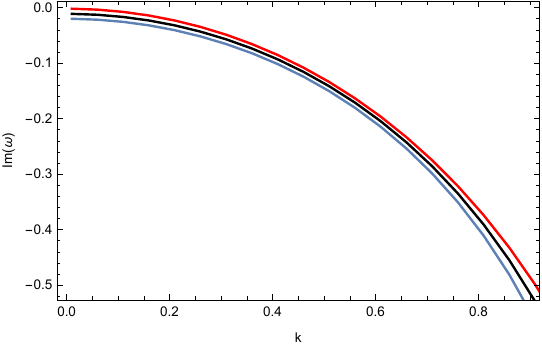}
	\end{subfigure}\hskip2cm
	\begin{subfigure}{0.35\textwidth}
		\includegraphics[width=1\linewidth]{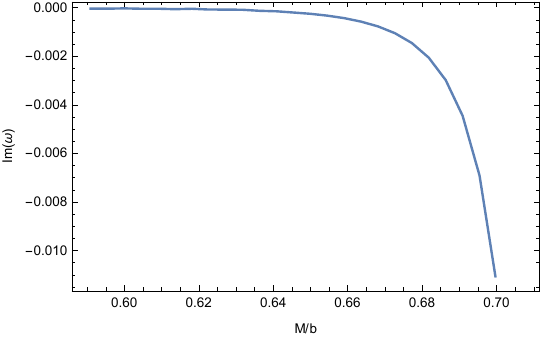}
	\end{subfigure}
	\caption{Left: Dispersion relation of the purely imaginary quasinormal mode in the axial longitudinal sector: near critical point for  M/b = 0.7045(Blue) and below the critical point at 0.7 (Black), 0.682 (Red). Right: Im$(\omega)$ vs $M/b$ near the critical point for k=0.001. We see that below the phase transition, the mode is, to a very good approximation, hydrodynamic, with only a tiny non zero value at zero momentum.}
	\label{diff-Mb}
\end{figure}

We have considered these gauge invariant quantities and studied the behaviour of the QNMs. In Fig. 	\ref{fig:long-qnm}, we have given the spectrum of these fields for the limit $M/b\rightarrow0$ with $k=0.1$. In this limit, the fluctuation of the scalar field ($\eta(u)$ and $\sigma(u)$) decouples from the two gauge fields ($a_t(u)$ and $a_z(u)$) due to the fact that the background scalar field vanishes at this limit. Thus the spectrum of the axial fields will be exactly as that of the vector field as shown in Fig. 	\ref{fig:long-qnm}. Here encircled blue dots corresponds to the spectrum of the axial field and the red circle correspond the spectrum of the vector field.  At this limit of $M/b\rightarrow0$ the mode can be considered as a diffusive mode. The remaining blue dots in Fig. \ref{fig:long-qnm}, corresponds to the modes which arises from the fluctuation of the scalar field. 

In Fig. \ref{long} we have plotted the Im$(w)$ vs $M/b$ for $k=0.001$ and $k=1$ for the mode which sits on the negative imaginary axis, where we can see that that the value of the mode remains constant up to the critical point i.e. around $M/b\approx 0.71$ where this diffusive mode starts turning into pseudo diffusive mode as could be expected in the case of symmetry breaking due to a scalar field \cite{Amado:2009ts,Jimenez-Alba:2015awa}. Above this critical limit this mode moves very fast to very negative imaginary values and thus does not represent hydrodynamic behavior anymore. We have presented the dispersion relation for this pseudo diffusive mode  in Fig.	\ref{diff-Mb}  where we can see a similar kind of dispersion as that of the $E_z (u)$ but at the limit of $k\rightarrow0$ this mode does not go exactly to zero. It is therefore a very long-lived pseudo diffusive mode which needs to be included in any hydrodynamic description\footnote{Inclusion of such modes is often denoted as "Hydro+" \cite{Stephanov:2017ghc}}. 


\section{Discussion}\label{sec:discussion}
We have studied the Hydrodynamics of a holographic model of Weyl semimetals by studying the behaviour of QNMs over the parameter space of the system. 
In addition to this we have also considered the variation of  coefficient of the pure Chern-Simon term $"\alpha"$ onto QNMs where we can see that with the increasing value of $\alpha$ the separation between the components of the perturbed fields increases linearly for a small value of $\alpha$. We have also studied the QNMs in the broken phase by considering the perturbation on the axial field in the longitudinal direction, there one can see that the system will produce a diffusive mode in the   non-trivial topological phase, around the critical point $M/b\approx 0.71$ this will turn into a pseudo-diffusive mode and in the trivial phase no such mode exist. 

\bibliographystyle{unsrt}
\bibliography{ref}
\end{document}